# Adjustment control usability and dynamic accommodation in office chairs: A comparative study of two contemporary designs

Yejin Kang[a, b], Dusan Baek[b], Dongwoo Park[b], Kitae Hwang[b], Woojin Park[b, c, *]

[a] Industrial and Operations Engineering, University of Michigan, Ann Arbor, United States

[b] Department of Industrial Engineering, Seoul National University, Seoul, Republic of Korea

[c] Institute for Industrial Systems Innovation, Seoul National University

[*] Corresponding author.

Department of Industrial Engineering, Seoul National University, 1 Gwanak-ro, Gwanak-gu, Seoul, 08826, South Korea

Tel: +82 (02) 880 4310

E-mail address: woojinpark@snu.ac.kr

**Abstract:**

Adjustability is a core determinant of ergonomic fit in office chairs, but its benefits depend on users being able to operate the controls and, on the chair, continuing to support the body as posture changes during work. We compared two high-end office chairs, the T90 (Sidiz; Chair A) and the Aeron (Herman Miller; Chair B), in two experiments to explore which design features may explain differences between them. In Experiment 1, 20 participants adjusted ten controls common to both chairs and rated perceived adjustment difficulty on four items. Chair A had significantly lower overall difficulty (Wilcoxon signed-rank test, $p = 0.008$), with control-level advantages for headrest angle, armrest height, armrest rotation angle and lumbar support position; Chair B was easier for seat height, possibly because its side-mounted lever is more visible than Chair A's underside button. The headrest comparisons involved a third-party headrest fitted to Chair B, and control- and item-level tests were not corrected for multiple comparisons. In Experiment 2, the same participants completed a 15-minute multi-device task requiring repeated posture changes, reach and reorientation, then rated body-part discomfort, general comfort and satisfaction with five components. Discomfort and comfort did not differ; Chair A had higher satisfaction for the seat-pan and lumbar support (exploratory, uncorrected). Interviews suggested links to mechanism properties: Chair B's rear lumbar dial was often described as producing little perceptible change, and for controls without significant differences each chair appeared to leave a different aspect of the same need unmet. These findings suggest mechanical and interface properties that designers of adjustable office chairs may consider for dynamic, multi-device work.

## 1. Introduction

Office chairs must accommodate variability arising from both users and their work. Users differ in anthropometric dimensions, physical characteristics, and sitting preferences, requiring chairs to accommodate substantial between-user variability (Kim et al., 2016; Yuan et al., 2022). At the same time, office work increasingly involves interaction with multiple devices and work surfaces, including desktop and laptop computers, smartphones, and tablets (Grudin, 2001; Jokela et al., 2015; Yuan et al., 2022). As workers move among these devices and work surfaces, they may repeatedly change their posture and body orientation and reach or move toward different locations. Thus, an office chair must accommodate not only differences between users but also frequent changes within the same user over the course of work.

Such accommodation can occur in two complementary ways, termed here adjustability and dynamic accommodation. Adjustability refers to the ability of users to deliberately modify chair settings according to their body characteristics, preferences, or work requirements. Dynamic accommodation refers to the chair's capacity to accommodate changes in users' posture and movement during work without requiring deliberate readjustment. It can involve the deformation or movement of chair components in response to users' body movements and applied forces, allowing the chair to conform to or follow them as their posture and position change. Thus, whereas adjustability enables users to configure a chair as needed, dynamic accommodation enables the chair to continue supporting their changing posture and movement during work. Despite the importance of both capabilities, important gaps remain in the evidence available to guide their design.

The first research gap concerns the design of adjustment controls. Considerable ergonomic knowledge is available concerning what aspects of an office chair should be adjustable and by how much. Previous research and guidelines have addressed anthropometric fit and adjustment requirements for features such as seat height, seat depth, seat-pan angle, backrest angle, lumbar support, and armrests (Allie & Kokot, 2005; BIFMA, 2013; Goonetilleke & Feizhou, 2001). Appropriate adjustment has also been associated with reduced physical discomfort and musculoskeletal loading (Amick et al., 2003; van Niekerk et al., 2012). Yet these benefits depend on users being able to operate the controls effectively. Many office workers do not appropriately adjust their chairs (Grandjean et al., 1983; Ong et al., 1988; Verbeek, 1991), and difficulties in identifying, understanding, or operating adjustment mechanisms can discourage their use (Allie & Kokot, 2005; Helander, 2003; Helander & Zhang, 2001; Kleeman, 1991; Underwood & Sims, 2019). Compared with the extensive guidance on what should be adjustable and by how much, much less is known about how adjustment controls should be designed for easy use. Although some studies have examined chair-control operation and adjustment methods (Groenesteijn et al., 2009; Helander et al., 1995), evidence remains limited regarding which characteristics of adjustment controls make them easier or more difficult to operate.

The second research gap concerns dynamic accommodation. Previous office-chair research has examined the effects of different chair designs on outcomes such as posture, movement, comfort, and physical loading (Gregory et al., 2006; Kingma & van Dieën, 2009; O'Sullivan et al., 2012; van Dieën et al., 2001; Vergara & Page, 2002), and some studies have specifically investigated chairs or mechanisms intended to permit or promote movement during sitting (Grooten et al., 2017; O'Sullivan et al., 2012). However, limited evidence is available on how well office chairs designed to provide dynamic accommodation actually accommodate users as their posture and movement change during work, or on which design characteristics may contribute to such accommodation. This knowledge is particularly relevant to contemporary office work, in which users may repeatedly change posture, reach, and reorient themselves while interacting with different devices and work surfaces.

To help fill these two research gaps, the present study comparatively evaluated two contemporary high-end office chairs, the T90 (Sidiz, Pyeongtaek, Republic of Korea) and the Aeron (Herman Miller, Zeeland, MI, USA). Both chairs provide extensive adjustability and incorporate features that accommodate users' changing posture and movement during work, but they do so through substantially different design approaches. This contrast provided an opportunity to examine both adjustment-control usability and dynamic accommodation. Two complementary experiments were conducted. Experiment 1 examined adjustment-control usability by comparing the perceived difficulty of operating individual controls and identifying control-design characteristics associated with differences in adjustment difficulty. Experiment 2 evaluated dynamic accommodation through users' experiences with the two chairs during a multi-device task involving repeated changes in posture, orientation, and reach. Together, the experiments provide empirical evidence on design characteristics relevant to effective adjustment and dynamic accommodation in contemporary office-chair design.

## 2. Methods

2.1. Subjects

Twenty participants (10 male and 10 female) without current neck or back pain participated in the study. Their mean age was 30.6 years (SD = 9.6), mean height was 1.69 m (SD = 0.10), and mean weight was 63.5 kg (SD = 11.4). The sample included six office workers, five graduate students, and nine undergraduate students. To minimize potential bias associated with prior brand familiarity, only individuals who reported being unfamiliar with both Herman Miller and Sidiz were recruited.

2.2. Office chairs

Two office chairs, the T90 (Sidiz) and a modified Aeron (Herman Miller), were selected for the study and are referred to hereafter as Chair A and Chair B, respectively. Chair B was modified by attaching a third-party headrest of similar material because the Aeron did not originally include a headrest. This modification provided Chair B with headrest height and angle adjustment functions, thereby enabling comparison with the corresponding functions of Chair A, although the adjustment mechanisms differed between the two chairs.

The two chairs were selected as informative contrasting cases for examining adjustability and dynamic accommodation. Both are contemporary high-end office chairs that provide extensive adjustability and incorporate design features that accommodate changes in users' posture and movement during sitting. However, they differ substantially in how these two capabilities are implemented: their adjustment functions use different control mechanisms, and their designs differ in how they respond to or accommodate users' postural and movement changes. These contrasts made the chairs suitable for examining the two design issues identified in the Introduction: the usability of adjustment controls and the design characteristics underlying dynamic accommodation.

With the addition of the headrest to Chair B, the two chairs shared ten adjustment functions, although the mechanisms used to implement these functions differed. Chair A additionally provided seat-depth adjustment, resulting in 11 adjustment functions for Chair A and 10 for Chair B. Table 1 compares the adjustment mechanisms of the two chairs, while Figures 1 and 2 show the locations of their adjustment controls. Appendix A provides images of the individual controls reproduced from the manufacturers' official online user manuals.

Table 1. Comparison of adjustment mechanisms of the two chairs

| ID | Adjustment dimension | Chair A | Chair B |
|---|---|---|---|
| F1 | Seat height | Seat height is adjusted using a pneumatic gas-lift mechanism operated by a button located on the right underside of the seat-pan. | Seat height is adjusted using a pneumatic gas-lift mechanism operated by a lever located on the right side of the seat-pan. |
| F2 | Seat depth | Seat depth is adjusted by sliding the seat-pan through six discrete positions. | Seat depth is not adjustable; instead, the chair is available in three fixed frame sizes. |
| F3 | Seat-pan tilt | Forward seat-pan tilt is engaged using a lever located on the left underside of the seat-pan. | Forward seat-pan tilt is engaged by leaning back against the backrest, operating a lever on the left underside of the seat-pan, shifting body weight forward, and locking the backrest in position. |
| F4 | Headrest height | Headrest height is adjusted by directly grasping and sliding the integrated headrest upward or downward. | The third-party headrest attached for this study is adjusted by loosening a screw, sliding the headrest upward or downward, and retightening the screw. |
| F5 | Headrest angle | Headrest angle is adjusted by directly grasping and tilting the head rest to the desired position. | Headrest angle is adjusted by loosen ing a screw, tilting the third-party headrest to the desired position, and |
| | | | retightening the screw. |
| F6 | Armrest height | Armrest height is adjusted using a button located near the armrest pad. The user presses the button, repositions the armrest, and releases the button to lock it in place. | Armrest height is adjusted using a l ever located behind the armrest. The user releases the lever, repositions the armrest, and re-locks the lever. |
| F7 | Armrest rotation angle | The upper and lower sections of the armrest can be rotated independently using the armrest's one-touch mechanism. | The armrest rotation angle is adjusted through three discrete positions. |
| F8 | Armrest fore-aft position | The fore-aft position of the armrest is adjusted through five discrete positions. | The fore-aft position of the armrest is adjusted continuously by sliding the armrest pad forward or backward. |
| F9 | Backrest recline tension | Backrest recline tension is adjusted using a quick-tension dial requiring approximately 2.5 rotations across its adjustment. | Backrest recline tension is adjusted using a dial requiring approximately 12.5 rotations across its adjustment. |
| F10 | Backrest recline limit | The maximum backrest recline is set using a four-position limiter. | The maximum backrest recline is set using a three-position limiter corresponding to upright, intermediate, and maximum recline s ettings. |
| F11 | Lumbar support position | The vertical position of the lumbar support is adjusted by grasping the levers at both sides of the backrest and sliding them upward or downward. | The fore-aft position of the lumbar support pad is adjusted by rotating a control located at the rear of the backrest. |

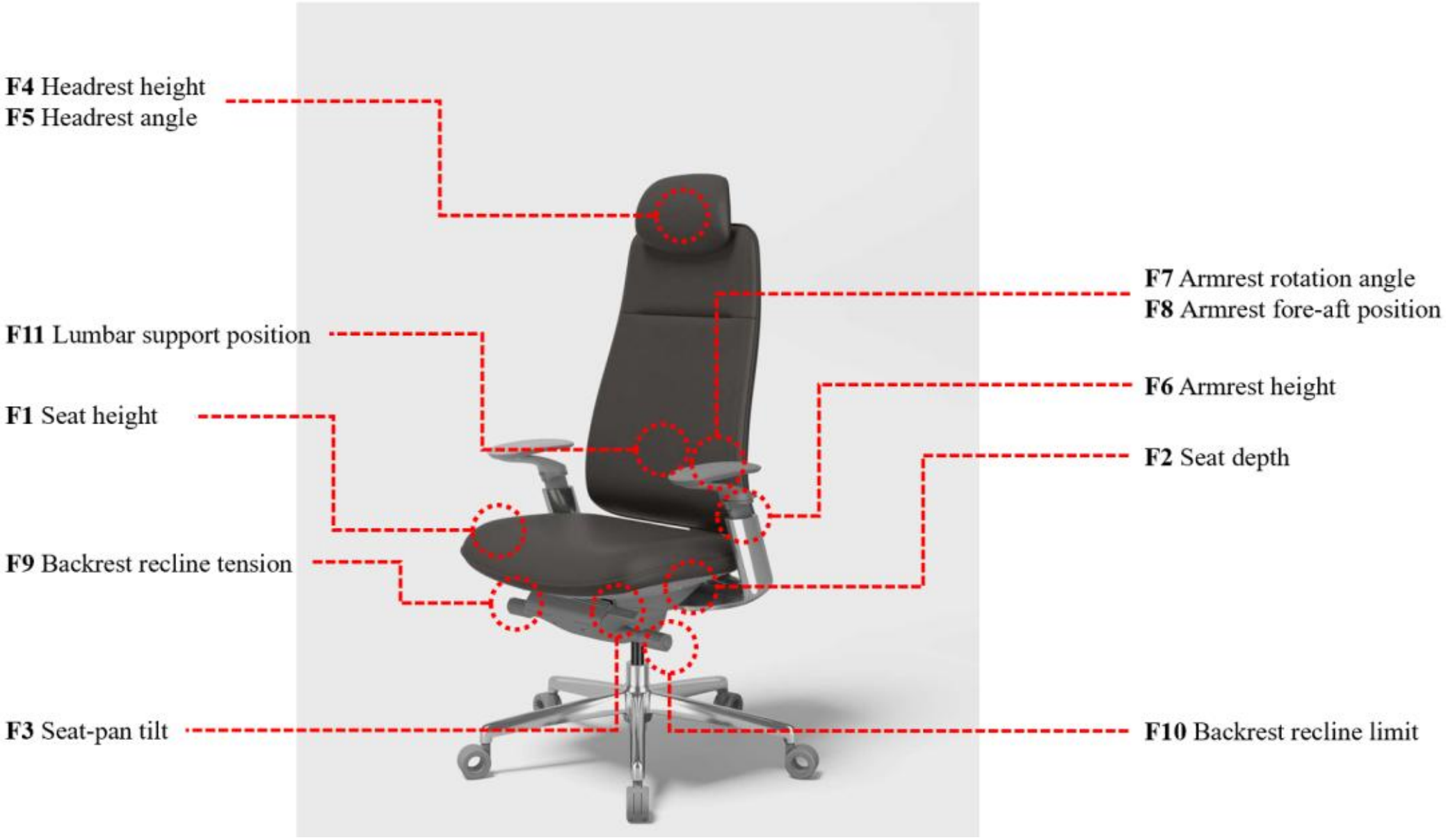

Figure 1. Locations of the adjustment controls for Chair A

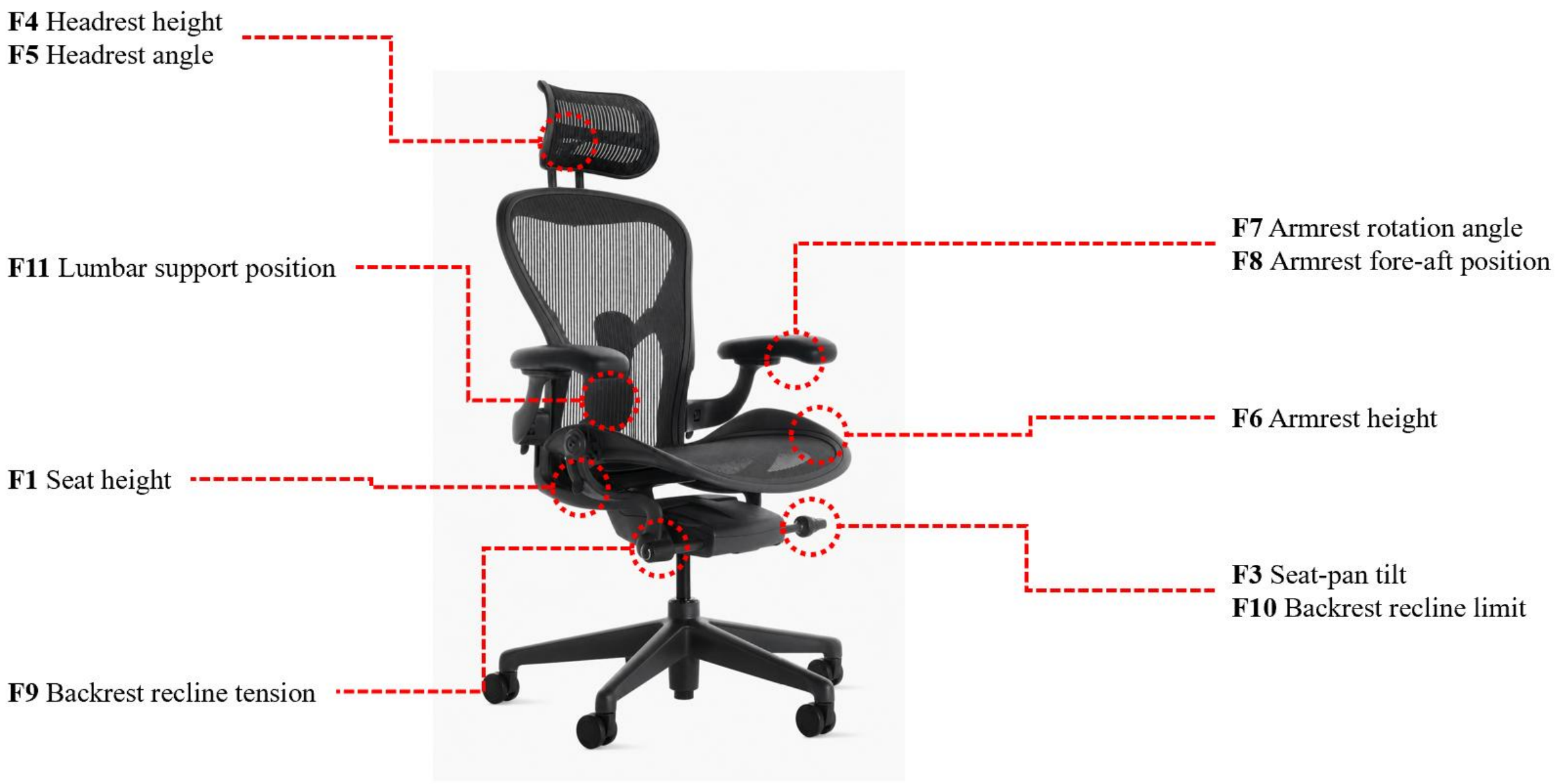

Figure 2. Locations of the adjustment controls for Chair B

### 2.3. Experiment 1: Adjustment-control usability

The study comprised two experiments corresponding to the two design issues identified in the Introduction. Experiment 1 evaluated the usability of the adjustment controls, whereas Experiment 2 examined dynamic accommodation during a movement-based task. All participants completed Experiment 1 before Experiment 2. The order of Chair A and Chair B was counterbalanced across participants in Experiment 1, and the order was reversed for each participant in Experiment 2.

#### 2.3.1. Procedure

Before each chair trial, all adjustable dimensions were set to their minimum settings or positions within their respective adjustment ranges, producing the most compact chair configuration. Participants then adjusted the dimensions one at a time in a fixed order corresponding to F1–F11 in Appendix A: seat height, seat depth, seat-pan tilt, headrest height, headrest angle, armrest height, armrest rotation angle, armrest fore-aft position, backrest recline tension, backrest recline limit, and lumbar support position. This order was based on prior observation of the sequence in which users most frequently adjusted the chair features. Because Chair B did not provide seat-depth adjustment, F2 was omitted for Chair B. The chair order was counterbalanced across participants.

For each adjustment dimension, the experimenter first identified the relevant control and its location without explaining how it was operated. The participant was then given unlimited time to determine the operating method independently. Once the participant either indicated that the method had been determined or reported being unable to determine it, the experimenter explained the correct operating method so that any initial misunderstanding was corrected. The participant then adjusted the dimension to the preferred setting and immediately rated the difficulty of operating the control. This sequence was repeated for each applicable adjustment dimension. After completing all adjustments and ratings for a chair, the participant was allowed to readjust any setting as needed to achieve their preferred overall configuration. The same procedure was then repeated with the second chair. The final settings for each chair, including any readjustments, were retained for Experiment 2.

#### 2.3.2. Measurement

Perceived difficulty of operating each adjustment control was assessed using four items adapted from the cognitive barriers to chair adjustment described by Underwood and Sims (2019). Participants responded to each item on a 7-point Likert scale ranging from 1 (strongly disagree) to 7 (strongly agree). Higher scores indicated greater perceived difficulty. The items were:

Q1. It was difficult to discover how to adjust the control.

Q2. The adjustment method of the control was too technical or complicated.

Q3. The experimenter's instructions for adjusting the control were difficult to understand.

Q4. It was difficult to adjust the control.

#### 2.3.3. Statistical analysis

Three levels of adjustment-difficulty scores were calculated. An item score was the response to an individual questionnaire item for a given adjustment function. An adjustment-dimension score was the mean of the four item scores for that dimension. An overall adjustment-difficulty score was the mean of the adjustment-dimension scores across the ten dimensions common to both chairs. Seat depth (F2), which was available only on Chair A, was therefore excluded from between-chair comparisons. Wilcoxon signed-rank tests were used to compare Chair A and Chair B at the item, adjustment-function, and overall levels. The significance level was set at $p \leq 0.05$, with no correction for multiple comparisons (see Section 4.3).

### 2.4. Experiment 2: Dynamic accommodation during movement

Experiment 2 examined participants' experience of the chairs while performing a task designed to induce repeated changes in posture, reach, and body orientation. The individualized chair settings established during Experiment 1 were retained for Experiment 2, allowing the chairs to be evaluated during movement under each participant's preferred configuration.

#### 2.4.1. Procedure

Each participant completed a 15-min movement-based task with each chair. As noted above, chair order was reversed from that used for the participant in Experiment 1.

The experimental setup consisted of two workstations positioned in opposite directions from the chair (Figure 3). The front workstation included a height-adjustable desk with two monitors and a smartphone placed at the right-hand corner of the desk. At a second workstation behind the chair, a human sculpture represented another person.

The two monitors displayed slightly different versions of the same passage. Participants continuously compared the passages, identified differing words, and corrected the version displayed on the right monitor. After each correction, they picked up the smartphone, entered the corrected word, and returned the phone to the desk, requiring repeated changes in posture and reach. Every three minutes, participants rotated 180° to face the workstation behind them, verbally reported the words corrected since the previous report, and then rotated back to resume the monitor task. This cycle continued for 15 min with each chair.

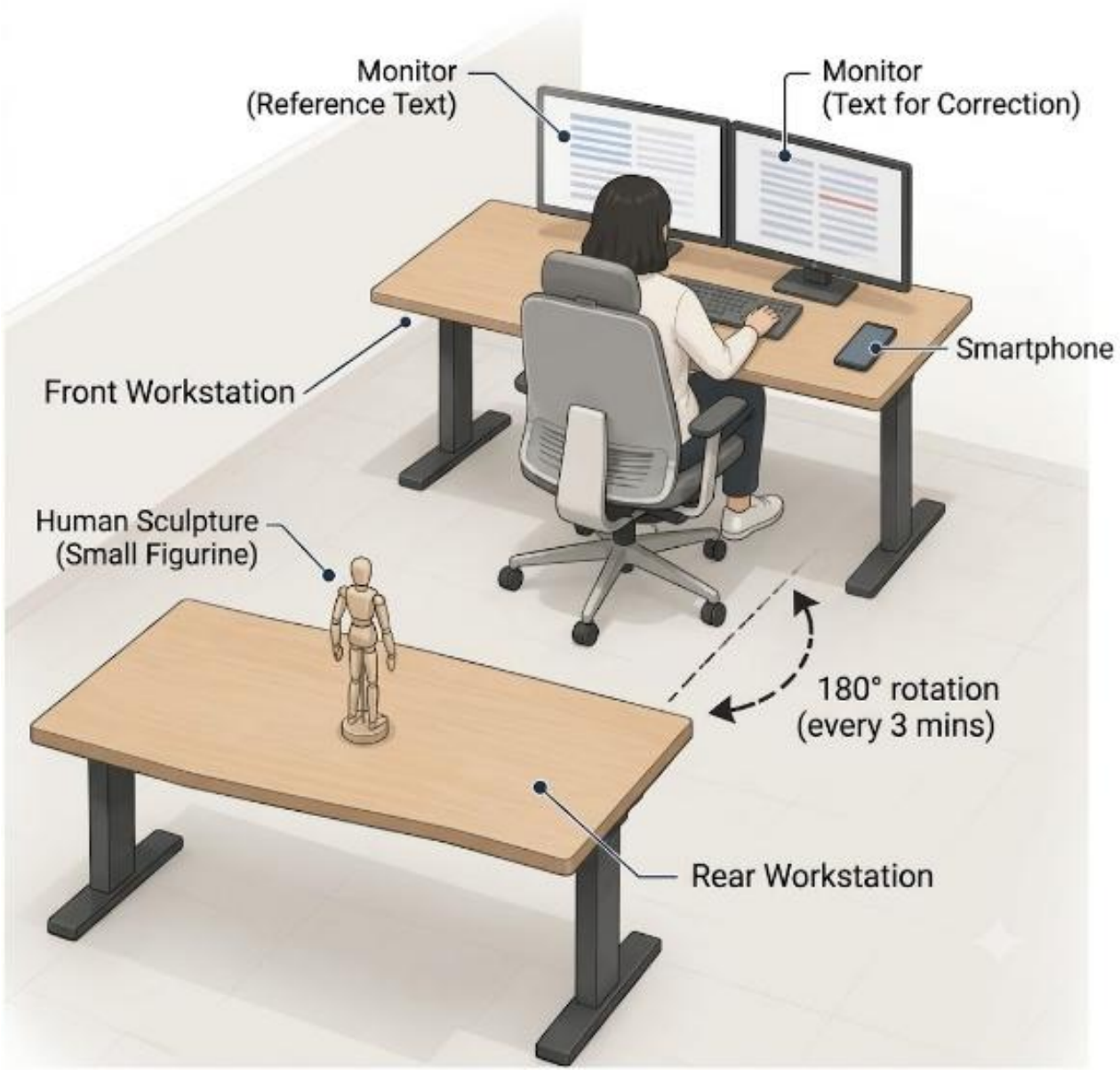

Figure 3. Experimental setup for Experiment 2

2.4.2. Measurement

Immediately after completing the 15-min task with each chair, participants rated body-part discomfort, overall sitting comfort, overall chair satisfaction, and satisfaction with five chair components.

Body-part discomfort was assessed for 12 body regions—neck, shoulders, upper arms, lower arms, upper back, mid back, lower back, buttocks, left thigh, right thigh, left calf, right calf—using the Body Part Discomfort scale (Corlett & Bishop, 1976). Ratings ranged from 0 to 5, with higher scores indicating greater discomfort. Overall sitting comfort was assessed using the General Comfort scale (Shackel et al., 1969), ranging from 0 (I feel unbearable pain) to 10 (I feel completely relaxed), with higher scores indicating greater comfort.

Overall chair satisfaction was rated on a 5-point scale, with higher scores indicating greater satisfaction (Channak et al., 2024). Participants also rated satisfaction with five chair components—seat-pan, headrest, armrest, backrest recline, and lumbar support—on a 5-point scale ranging from 1 (very dissatisfied) to 5 (very satisfied). When a participant rated any component 3 or lower, the experimenter asked the participant to explain the reason for the rating.

2.4.3. Statistical analysis

Body-part discomfort scores for each of the 12 body regions were compared between Chair A and Chair B using Wilcoxon signed-rank tests. Overall sitting comfort, overall chair satisfaction, and satisfaction with each of the five chair components were compared between chairs using the same test. The significance level was set at $p \leq 0.05$, with no correction for multiple comparisons (see Section 4.3). Responses to the follow-up questions were grouped into recurring categories based on similarity of content.

## 3. Results

3.1. Experiment 1: Adjustment-control usability

For the overall adjustment score, aggregated across all subjects and the ten controls common to both chairs, Chair A received a significantly lower (i.e., easier-to-adjust) score than Chair B (Chair A: $M = 1.88$, $SD = 0.54$; Chair B: $M = 2.30$, $SD = 0.76$; Wilcoxon signed-rank test, $p = 0.008$; Table 2), indicating that participants found Chair A easier to adjust overall.

At the control level, five of the ten common controls showed a significant difference between chairs (significant data in Table 3). For seat height (F1), Chair A received a significantly higher (less favorable) score than Chair B (Chair A: M = 2.62, SD = 1.14; Chair B: M = 2.08, SD = 1.31; p = 0.017), indicating that Chair B's seat height control was easier to adjust. In contrast, Chair A was rated significantly easier to adjust than Chair B for headrest angle (F5; Chair A: M = 1.05, SD = 0.10; Chair B: M = 1.96, SD = 1.43; p = 0.005), armrest height (F6; Chair A: M = 1.38, SD = 0.76; Chair B: M = 2.29, SD = 1.08; p = 0.001), armrest rotation angle (F7; Chair A: M = 1.36, SD = 0.53; Chair B: M = 2.22, SD = 1.34; p = 0.014), and lumbar support position (F11; Chair A: M = 1.54, SD = 0.77; Chair B: M = 1.94, SD = 1.05; p = 0.015). No significant differences between chairs were found for the remaining five common controls—seat-pan tilt (F3), headrest height (F4), armrest fore-aft position (F8), backrest recline tension (F9), and backrest recline limit (F10).

At the item level, 15 of the 40 items (across the ten common controls) showed significant differences between chairs (significant data in Table 4). For seat height (F1), participants found it significantly more difficult to discover how to adjust the control for Chair A than for Chair B (Chair A: M = 4.90, SD = 1.92; Chair B: M = 3.40, SD = 2.30; p = 0.019), and rated Chair A's adjustment method as significantly more technical or complicated than Chair B's (Chair A: M = 2.10, SD = 1.37; Chair B: M = 1.80, SD = 1.24; p = 0.034); both items favored Chair B, consistent with the control-level result for F1. For seat-pan tilt (F3), although the control-level score did not differ significantly, participants rated Chair A's adjustment method as significantly less technical or complicated than Chair B's (Chair A: M = 2.35, SD = 1.57; Chair B: M = 3.65, SD = 1.93; p = 0.027).

For headrest height (F4), Chair A's adjustment method was rated as significantly less technical or complicated than Chair B's (Chair A: M = 1.10, SD = 0.31; Chair B: M = 2.20, SD = 1.58; p = 0.011), despite no significant difference at the control level. For headrest angle (F5), all three significant items favored Chair A: it was significantly less difficult to discover how to adjust (Chair A: M = 1.10, SD = 0.31; Chair B: M = 1.85, SD = 1.53; p = 0.041), the adjustment method was significantly less technical or complicated (Chair A: M = 1.00, SD = 0.00; Chair B: M = 2.00, SD = 1.72; p = 0.016), and the control was significantly less difficult to adjust overall (Chair A: M = 1.10, SD = 0.31; Chair B: M = 2.45, SD = 2.14; p = 0.018), consistent with the control-level result.

For armrest height (F6), all three significant items favored Chair A: it was significantly less difficult to discover how to adjust (Chair A: M = 1.50, SD = 0.95; Chair B: M = 3.20, SD = 1.99; p = 0.005), the adjustment method was significantly less technical or complicated (Chair A: M = 1.25, SD = 0.55; Chair B: M = 2.10, SD = 1.48; p = 0.019), and the control was significantly less difficult to adjust overall (Chair A: M = 1.40, SD = 0.88; Chair B: M = 2.55, SD = 1.79; p = 0.007). For armrest rotation angle (F7), the two significant items likewise favored Chair A: it was significantly less difficult to discover how to adjust (Chair A: M = 1.55, SD = 1.15; Chair B: M = 3.00, SD = 2.15; p = 0.010), and the control was significantly less difficult to adjust overall (Chair A: M = 1.50, SD = 1.15; Chair B: M = 2.45, SD = 1.73; p = 0.044). For armrest fore-aft position (F8), although the control-level score did not differ significantly, it was significantly more difficult to discover how to adjust this control for Chair A than for Chair B (Chair A: M = 2.05, SD = 1.90; Chair B: M = 1.10, SD = 0.31; p = 0.045), favoring Chair B.

Finally, for lumbar support position (F11), the two significant items both favored Chair A: the instructions for adjusting the control were significantly less difficult to understand (Chair A: M = 1.25, SD = 0.55; Chair B: M = 1.70, SD = 1.26; p = 0.039), and the control was significantly less difficult to adjust overall (Chair A: M = 1.45, SD = 1.00; Chair B: M = 2.40, SD = 1.79; p = 0.028), consistent with the control-level result. No items reached significance for seat depth (F2, unique to Chair A), backrest recline tension (F9), or backrest recline limit (F10).

Table 2. Results for overall score

| **Chair A** | | **Chair B** | | **p-value** | **Pairwise Comparison Result** |
|---|---|---|---|---|---|
| **Mean** | **SD** | **Mean** | **SD** | | |
| 1.88 | 0.54 | 2.3 | 0.76 | 0.008 | Chair A (Better) < Chair B |

Table 3. Results for control score

| **Control** | **Chair A** | | **Chair B** | | **p-value** | **Pairwise Comparison Result** |
|---|---|---|---|---|---|---|
| | **Mean** | **SD** | **Mean** | **SD** | | |
| F1 | 2.62 | 1.14 | 2.08 | 1.31 | 0.017 | Chair A > Chair B (Better) |
| F5 | 1.05 | 0.1 | 1.96 | 1.43 | 0.005 | Chair A (Better) < Chair B |
| F6 | 1.38 | 0.76 | 2.29 | 1.08 | 0.001 | Chair A (Better) < Chair B |
| F7 | 1.36 | 0.53 | 2.22 | 1.34 | 0.014 | Chair A (Better) < Chair B |
| F11 | 1.54 | 0.77 | 1.94 | 1.05 | 0.015 | Chair A (Better) < Chair B |

Table 4. Results for item score

| **Control** | **Item** | **Chair A** | | **Chair B** | | **p-value** | **Pairwise Comparison Result** |
|---|---|---|---|---|---|---|---|
| | | **Mean** | **SD** | **Mean** | **SD** | | |
| F1 | Q1 | 4.90 | 1.92 | 3.40 | 2.30 | 0.019 | Chair A > Chair B (Better) |
| | Q2 | 2.10 | 1.37 | 1.80 | 1.24 | 0.034 | Chair A > Chair B (Better) |
| F3 | Q2 | 2.35 | 1.57 | 3.65 | 1.93 | 0.027 | Chair A (Better) < Chair B |
| F4 | Q2 | 1.10 | 0.31 | 2.20 | 1.58 | 0.011 | Chair A (Better) < Chair B |
| F5 | Q1 | 1.10 | 0.31 | 1.85 | 1.53 | 0.041 | Chair A (Better) < Chair B |
| | Q2 | 1.00 | 0.00 | 2.00 | 1.72 | 0.016 | Chair A (Better) < Chair B |
| | Q4 | 1.10 | 0.31 | 2.45 | 2.14 | 0.018 | Chair A (Better) < Chair B |
| F6 | Q1 | 1.50 | 0.95 | 3.20 | 1.99 | 0.005 | Chair A (Better) < Chair B |
| | Q2 | 1.25 | 0.55 | 2.10 | 1.48 | 0.019 | Chair A (Better) < Chair B |
| | Q4 | 1.40 | 0.88 | 2.55 | 1.79 | 0.007 | Chair A (Better) < Chair B |
| F7 | Q1 | 1.55 | 1.15 | 3.00 | 2.15 | 0.01 | Chair A (Better) < Chair B |
| | Q4 | 1.50 | 1.15 | 2.45 | 1.73 | 0.044 | Chair A (Better) < Chair B |
| F8 | Q1 | 2.05 | 1.90 | 1.10 | 0.31 | 0.045 | Chair A > Chair B (Better) |
| F11 | Q3 | 1.25 | 0.55 | 1.70 | 1.26 | 0.039 | Chair A (Better) < Chair B |
| | Q4 | 1.45 | 1.00 | 2.40 | 1.79 | 0.028 | Chair A (Better) < Chair B |

3.2. Experiment 2: Dynamic accommodation during movement

3.2.1. Subjective ratings

For body part discomfort, no significant differences between chairs were found for any of the twelve body regions (all $p > 0.05$). For general sitting comfort, Chair A received a numerically higher (more comfortable) rating than Chair B, though the difference was not statistically significant (Chair A: M = 7.65, SD = 1.04; Chair B: M = 7.15, SD = 1.95; $p = 0.387$).

Overall satisfaction with the chair was numerically higher for Chair A than for Chair B, but the difference did not reach significance (Chair A: M = 4.20, SD = 0.52; Chair B: M = 3.85, SD = 0.88; $p = 0.071$). For satisfaction with the five components, participants were significantly more satisfied with Chair A's seat-pan than with Chair B's (Chair A: M = 4.15, SD = 0.49; Chair B: M = 3.55, SD = 1.10; $p = 0.030$) and significantly more satisfied with Chair A's lumbar support than with Chair B's (Chair A: M = 3.85, SD = 0.88; Chair B: M = 3.20, SD = 1.01; $p = 0.019$) (Table 5). No significant differences were found for the headrest (Chair A: M = 3.50, SD = 1.00; Chair B: M = 3.75, SD = 0.85; $p = 0.236$), the armrest (Chair A: M = 3.95, SD = 1.10; Chair B: M = 4.10, SD = 0.79; $p = 0.512$), or the backrest recline range and angle (Chair A: M = 3.85, SD = 0.99; Chair B: M = 3.95, SD = 0.83; $p = 0.642$); satisfaction was numerically slightly higher for Chair B than for Chair A on these three components, though not significantly so.

Table 5. Results for satisfaction with each of five chair components

| **Measure** | **Chair A** | | **Chair B** | | **p-value** | **Pairwise Comparison Result** |
|---|---|---|---|---|---|---|
| | **Mean** | **SD** | **Mean** | **SD** | | |
| Seat-pan | 4.15 | 0.49 | 3.55 | 1.10 | 0.03 | Chair A (Better) > Chair B |
| Lumbar support | 3.85 | 0.88 | 3.20 | 1.01 | 0.019 | Chair A (Better) > Chair B |

Responses on interviews were grouped into themes by control cluster with authors' decisions (Table 6). N is the number of participants who raised this theme, out of those interviewed about that control on that chair (interviews were triggered only for controls rated ≤ 3 on the 1-5 satisfaction scale). Representative comments by participants are written in Table 6. For Chair A, headrest was the most frequently discussed control (7 of 20 participants), with the dominant theme being that the headrest simply did not engage in a normal working posture.

It only made contact when leaning back significantly, leaving most participants without a strong opinion either way. Backrest recline and lumbar support each drew five comments, centered on the recline lacking an adjustable middle position and the lumbar adjustment producing little perceptible change. Only one participant raised a concern about the seat-pan.

For Chair B, the seat-pan and lumbar support were by far the most frequently discussed controls. Seven participants commented on the seat-pan, most describing the surface as too hard and citing pressure discomfort, which are directly consistent with Chair B's significantly lower seat-pan satisfaction in Table 5. Lumbar support drew the most comments of any control cluster for either chair (12 of 20 participants), and the most frequent theme was that the adjustment had little effect. Half of those participants (6 of 12) described the lumbar adjustment as producing little or no perceptible change in support, with several noting they could not tell whether the feature was engaged at all. This pattern closely mirrors Chair B's significantly lower lumbar support satisfaction relative to Chair A (Table 5), suggesting that the adjustment mechanism's lack of a perceptible effect may have contributed to that difference, alongside comments about the firmness and material of the support itself. Comments on Chair B's headrest, armrest, and backrest recline were more mixed, ranging from positive (the headrest's angle adjustment was reported to noticeably improve support for one participant) to critical (stiff armrest adjustment requiring excessive force; insufficient upper-back coverage from the backrest), consistent with the absence of significant between-chair differences for these three components in Table 5.

Table 6. Interview responses on components

| Component | Chair | n | Theme | Representative comment |
|---|---|---|---|---|
| Seat-pan | A | 1 | Insufficient depth/ cushioning relative to preference | "Less cushioned than what I'm used to; even fully extended, the seat depth wasn't enough for me." (P17) |
| | B | 4 | Hard surface causing pressure discomfort | "The material is hard, so my hips felt sore/marked after sitting for a while." (P06) |
| | | 2 | Inadequate seat depth / sizing | "There's no depth adjustment, so the seat-pan felt short relative to my thigh length." (P14) |
| | | 2 | Strong contouring restricts posture freedom | "The curvature is strong, so my knee position gets fixed and I can't readjust comfortably." (P19) |
| | | 1 | Material preference (mesh vs. cushioned) | "I don't prefer mesh material; I'd rather have a cushioned one." (P14) |
| Headrest | A | 6 | Doesn't engage / no perceptible support in normal (non-reclined) posture | "Unless I lean back a lot, the headrest doesn't touch my head, so I didn't have a strong preference either way." (P07) |
| | | 1 | Awkward position causes neck strain when engaged | "It protrudes inward, so I have to push my head forward; to avoid it I end up straining my neck." (P06) |
| | B | 3 | Underuse / neutral, rarely leaned on | "I don't lean on headrests much even in my car; I don't really use it, so I'm not sure." (P02) |
| | | 1 | Angle adjustment improved support (positive) | "Changing the angle made it support me much better - it was good." (P09) |
| | | 1 | Wished for greater range of motion | "More comfortable than Chair A, but I wish the headrest itself could tilt further back past the backrest." (P06) |
| | | 1 | Structural gap / lack of fixation | "The gap between the low backrest and the headrest is large, and it's not fixed, so my neck isn't supported - only my mid-back is." (P19) |
| Armrest | A | 2 | Not fixed / moves around, distracting | "It's not fixed, so it kept shifting sideways when I rested my elbow - I kept noticing it." (P19) |
| | | 1 | Low task-relevance (rarely used due to posture) | "I barely used it since I was reaching forward for the task, so it was hard to evaluate." (P10) |
| | | 1 | Left-right asymmetry | "Only my left arm felt comfortable; the right side was different." (P02) |
| | B | 2 | Adjustment mechanism too stiff / requires excessive force | “Matching both sides to the same height was hard, and the angle adjustment needed a lot of force.” (P08) |
| | | 1 | Instability / play in the joint | “Not painful, but it felt unstable - there was some wobble/play.” (P19) |
| Backrest recline | A | 2 | Recline tension/range issues (reclines too easily, lacks a moderate middle position) | “There's no in-between - it's either too reclined or too upright; the position I wanted was hard to reach.” (P16) |
| | | 2 | Comfortable reclined but not in upright/working posture | “The depth changes from the bottom to the top; comfortable when leaning back, but slightly uncomfortable in a normal seated posture.” (P18) |
| | | 1 | Low task-relevance (rarely leaned on) | “I was leaning forward for the task, so I never really rested against it - hard to evaluate.” (P10) |
| | B | 1 | Overly sensitive; must hold position actively | “It reacts too sensitively - I wanted to move just a little, but it moves a lot; to hold a position I had to keep applying force.” (P14) |
| | | 1 | Lacks upper-back support/ coverage | “It supports my mid-back well, but there's nothing supporting my upper back, so it'll probably get stiff over long use.” (P19) |
| | | 1 | General discomfort (unspecified) | “The feel was less comfortable than before.” (P11) |
| Lumbar support | A | 2 | Adjustment has little perceptible effect | “Raising or lowering it didn't feel very different.” (P09) |
| | | 1 | Hard/irritating material | “It felt hard and irritating rather than actually supportive.” (P20) |
| | | 1 | Positive / no issue (clearly better than Chair B) | “Nothing unusual - and it's clearly more comfortable than Chair B.” (P18) |
| | | 1 | Low task-relevance | “Since I was leaning forward, I didn't feel like I needed lumbar support at all.” (P10) |
| | B | 6 | Adjustment has little/no perceptible effect (“can't tell it's working”) | “I couldn't tell whether the feature was even there - adjusting it didn't produce a noticeable change.” (P10) |
| | | 2 | Insufficient firmness/ support intensity | “I wish it supported my lower back more tightly, but I didn't feel that.” (P05) |
| | | 1 | Hard material, wants more cushioning | “It's hard because of the overall material, so I wish it had more cushioning.” (P06) |
| | | 1 | Doesn't track dynamic posture changes | “When I shifted my posture back and forth, the support didn't move with me, so it felt less supportive.” (P03) |
| | | 1 | Missing height adjustment (presenton Chair A) | “Chair A let me adjust it up and down, but this one doesn't - that was disappointing.” (P17) |
| | | 1 | Feels like it locks a single posture | “It felt like it forced me into one fixed posture.” (P18) |

## 4. Discussion

This study compared Chair A and Chair B along two complementary dimensions: how easily their controls could be operated (Experiment 1) and how well the chairs supported users during a dynamic, multi-device task (Experiment 2). Consistent with the study's aim of moving beyond product-level comparison toward design-level explanation, the pattern of significant and non-significant difference across both experiments, combined with the mechanism-level differences between the two chairs and participants' interview responses, points to specific interface features that may account for the two chairs' relative strengths and weaknesses.

4.1. Findings on controls with significant difference

Seat height was the only control for which Chair B was rated significantly easier to adjust than Chair A (Experiment 1: Table 3), a direction opposite to every other significant control-level finding in this study. Both chairs use the same underlying mechanism (a pneumatic gas-lift cylinder), so the difference cannot be attributed to differences in mechanical principle; instead, it plausibly lies in the operating interface (Table 1): Chair B's height control is a lever positioned on the side of the seat-pan, in a location and form that is directly visible and recognizable as an adjustable control, whereas Chair A's control is a button located on the underside of the seat-pan, out of the user's direct line of sight. This distinction maps closely onto the specific items that drove the item-level difference (Table 4): participants found it significantly harder to discover how to adjust Chair A's control and rated its adjustment method as significantly more technical or complicated, rather than reporting that the physical motion of raising or lowering the seat itself was harder to execute. This suggests that for at least one type of control, adjustment difficulty can arise primarily from the visibility and recognizability of a control's affordance rather than from the mechanical operation it performs once found. This distinction is not captured by comparing adjustment mechanisms (e.g., button vs. lever) in isolation from their placement.

For headrest angle, Chair A was rated significantly easier to adjust across all three significant items (Experiment 1: Table 3, Table 4), a difference well explained by a clear procedural distinction between the two chairs (Table 1). Chair A's integrated headrest is repositioned with a single direct action, grasping and tilting it to the desired angle, whereas the third-party headrest used with Chair B requires a multi-step procedure: loosening a screw, tilting the headrest to the desired position, and retightening the screw. This additional procedural burden, rather than a difference in the range or precision of adjustment, plausibly explains why Chair B was rated as significantly harder to discover how to adjust, more technical or complicated, and more difficult to adjust overall. Because Chair B's headrest was a third-party attachment added specifically for this study rather than a manufacturer-integrated feature, this result may partly reflect the limitations of that substitution rather than a general design principle for headrest-angle adjustment; nonetheless, it illustrates that adjustment procedures requiring an intermediate step such as loosening and retightening a fastener impose a meaningfully higher usability cost than direct manipulation, even when the physical motion of tilting itself is simple.

For lumbar support, Chair A was rated significantly easier to adjust (Experiment 1: Table 3, Table 4) and significantly more satisfying during the movement task (Experiment 2: Table 5) than Chair B. This convergent pattern across both experiments is well explained by the underlying mechanism difference between the two chairs. Chair A's lumbar support can be adjusted intuitively by gripping the levers on both sides of the backrest and sliding them up and down, whereas Chair B relies on a dial mounted at the rear of the backrest that is farther from the user's reach, operates by rotation rather than direct vertical movement, and offers no height adjustment at all. This design gap is consistent with the interview data (Table 6): the dominant theme among Chair B's lumbar support comments, raised by 6 of the 12 participants interviewed about this control, was that adjusting the feature produced no perceptible change in support, with several participants stating they could not tell whether the mechanism was engaged at all. One participant explicitly contrasted the two chairs, noting that Chair A allowed height adjustment "up and down" while Chair B did not, consistent with the mechanism-level explanation in Table 1. However, the same complaint appeared in two of the five comments on Chair A's lumbar support, so it was not unique to Chair B, and the two controls also adjust different parameters (vertical position for Chair A, fore-aft pad position for Chair B; Table 1). Together, the quantitative results from both experiments and the qualitative interview data are compatible with an explanation in which, rather than being positioned poorly in an absolute sense, Chair B's rear-mounted, rotation-only, height-fixed mechanism may fail to give users a way to feel and control the support it provides.

Seat-related satisfaction followed a similar pattern, though the underlying explanation is only partially captured by the chairs' documented design differences. Chair A's seat-pan received significantly higher satisfaction ratings than Chair B's (Table 5). Chair A provides seat-depth adjustment across six discrete positions, whereas Chair B offers no seat-depth adjustment at all and instead relies on a choice of three fixed frame sizes to accommodate different body sizes (Table 1), a substitutive strategy that could not be exercised in this study, since a single Aeron unit was tested. This difference maps closely onto one of the two recurring complaint themes for Chair B's seat-pan in the interview data: two participants explicitly described the seat as too short relative to their thigh length with no way to adjust it. However, the more frequently raised theme for Chair B's seat-pan (4 of 7 participants who commented) concerned the hardness of the seating surface and resulting pressure discomfort during prolonged sitting, a material and cushioning property that is not addressed by the adjustability comparison above. This suggests that Chair A's seat-pan advantage in Experiment 2 may reflect at least two independent factors: adjustable depth and possibly a softer, more cushioned surface, only the first of which is a documented mechanism difference (Table 1). The second remains an interview-supported but otherwise unverified explanation and would benefit from direct measurement (e.g., seat surface hardness or pressure mapping) in future work.

Two further significant findings from Experiment 1, armrest height (F6) and armrest rotation angle (F7), both easier to adjust on Chair A, are also consistent with documented mechanism differences (Table 1). Chair A's armrest height is adjusted via a button located near the armrest pad itself: the user presses the button, repositions the armrest, and releases the button to lock it in place, a low-effort, three-step operation. Chair B's armrest height, by contrast, is adjusted via a lever positioned behind the armrest that must be released, repositioned, and re-locked, a higher-effort sequence resembling the operation of a bicycle-seat clamp. For armrest rotation angle, Chair A's upper and lower sections rotate independently through a one-touch mechanism, whereas Chair B's armrest rotates through only three discrete positions; the greater range and independence of adjustment available on Chair A plausibly made it easier for participants to find and reach a comfortable angle. Together, these mechanism differences plausibly explain why participants found Chair A's armrest controls easier to locate and operate during Experiment 1. Notably, however, this advantage in ease of adjustment did not carry over into armrest satisfaction during the dynamic movement task in Experiment 2, where the two chairs did not differ significantly and Chair B was even rated numerically (non-significantly) higher. The interview data help explain this gap: two participants using Chair A during Experiment 2 reported that its armrest was not fixed and shifted position during use, which is a stability problem distinct from the ease of initially setting the armrest's height or angle. This points to a design-level distinction worth stating explicitly: an accessible, low-effort adjustment mechanism does not by itself guarantee that the resulting position stays fixed under the repeated reaching and reorientation that dynamic, multi-device work requires. Armrest design for dynamic use may need to satisfy both properties: easy to adjust and secure once set, rather than the first alone.

### 4.2. Findings on controls with no significant difference

Three components, headrest, armrest, and backrest recline, showed no significant between-chair difference in Experiment 2 satisfaction (Table 5), even though the chairs implement these controls very differently and, in the case of the armrest, Chair A held a significant ease-of-adjustment advantage in Experiment 1. In each case, the interview data (Table 6) suggest that neither chair fully met the underlying user need, with each leaving a different aspect unmet. However, with N = 20, limited statistical power is an equally plausible reason for the absence of significant differences, and these interpretations are post hoc.

For the headrest, survey ratings did not differ (Table 5) but the interview results suggest the headrest could be improved in both chairs to actually engage the head and neck during typical, non-reclined desk work rather than only during heavy recline. Currently, neither Chair A's nor Chair B's headrest contains the design content participants wanted in this respect: on Chair A, six of seven participants who commented on the headrest said it simply did not make contact with their head or neck unless they leaned back substantially, leaving it functionally unused during normal work postures. On Chair B, three participants described underuse, and one participant pointed to a structural gap between a low backrest and the headrest and a lack of fixation, so that the neck remained unsupported even when the headrest was in range. Because both headrests may fail to deliver support during ordinary upright work (one through insufficient reach/engagement, the other through a positioning and fixation gap), satisfaction ratings for the two chairs may have converged despite the underlying problems being different. This suggests headrests intended for dynamic, multi-device work should be designed to make and maintain head/neck contact across the range of postures used during actual tasks, not only when a user reclines fully.

For the armrest, survey ratings did not differ (Table 5), but the interview results suggest the armrest could be improved in both chairs by resolving a stability/effort trade-off that neither design has fully solved. Currently, Chair A's and Chair B's armrests each lack the design content participants wanted: Chair A's armrest, despite being the easier of the two to adjust (Experiment 1), was reported by two participants to move or shift during the task rather than staying fixed once positioned, while Chair B's armrest was reported by two participants to require excessive force to adjust, with a further participant describing it as unstable despite its stiffness. Because Chair A appeared to trade stability for adjustability and Chair B adjustability for stability (each based on comments from only two participants), neither chair appeared to give participants an armrest that was both easy to set and reliably fixed during dynamic reaching and reorientation, which may be why overall satisfaction did not differ between them. This suggests armrest mechanisms for dynamic work should aim to decouple these two properties, for example, a low-effort adjustment step (as in Chair A) paired with a firmer locking mechanism once released (unlike Chair A), rather than treating ease of adjustment and post-adjustment stability as a single trade-off.

For backrest recline, ratings did not differ significantly between the two chairs (Table 3, 5), a result that is, on its face, difficult to reconcile with the mechanism comparison, which credits Chair A with an advantage on two counts: more recline positions (a four-position limiter, versus a three-position limiter spanning upright, intermediate, and maximum recline settings for Chair B) and a substantially lower-effort tension adjustment (2.5 quick-tension rotations to reach the desired strength, versus 12.5 rotations for Chair B; Table 1). Despite this documented mechanical advantage, the interview data suggest that participants had difficulty obtaining a stable middle range of recline that remained comfortable in an upright working posture, not only when fully reclined, on either chair. Because both chairs allow users to adjust recline tension and range (Table 1), however, these comments should not be taken to indicate that either chair lacks the necessary function. The individualized settings chosen in Experiment 1 were retained in Experiment 2 (Section 2.4), so the comments may instead reflect how those settings, selected during a brief adjustment-focused trial, were experienced during the dynamic, multi-device task. On Chair A, participants described the recline as lacking a moderate middle position (either "too reclined or too upright") and as comfortable only when leaning back, becoming less so in a normal seated posture. This may reflect the tension and range settings in use rather than the absence of an intermediate option. On Chair B, participants described the opposite failure mode: the mechanism was "too sensitive," requiring active, sustained force just to hold a chosen position, and it left the upper back unsupported even when the mid-back was well supported. Here too, the reported sensitivity may partly reflect the tension setting chosen in Experiment 1. Chair A's advantage in step count and tension effort therefore did not translate into a comfort advantage during the task, possibly because the specific complaints on each chair (lack of a stable middle position for Chair A, excessive sensitivity and incomplete back coverage for Chair B) related to the settings in use during dynamic work and to back coverage rather than to the dimensions the recline-mechanism comparison addresses. This suggests that the number of recline positions and the physical effort required to adjust tension, while relevant to Experiment 1's ease-of-adjustment findings for other controls, are not sufficient by themselves to produce a backrest that remains comfortable across the range of postures dynamic work requires; users may need support in selecting tension and range settings suited to dynamic work, and recline mechanisms may also need to maintain consistent upper- and lower-back contact throughout the adjustable range, not only at its endpoints.

### 4.3. Limitations and future directions

These findings should be interpreted in light of the study's scope. The comparison involved two specific commercial chairs rather than a systematic manipulation of individual mechanism features, so the design-level explanations offered above are necessarily inferential rather than causally isolated; a chair that implemented, for example, Chair A's lumbar-lever design on an otherwise Chair B-like frame would be needed to confirm that this feature in isolation drives the observed difference. The sample (N = 20) also limits statistical power, particularly for controls where between-chair differences were smaller or more variable. Finally, because interviews were only conducted following Experiment 2, the design-level explanations offered for Experiment 1's findings on seat height and headrest angle rely on the mechanism comparison alone and could not be corroborated with participant reasoning. In addition, the individualized settings chosen in Experiment 1 were retained in Experiment 2, so some interview comments, such as those on backrest recline, may reflect the settings chosen at that time rather than limits of the mechanisms themselves. Future work could address these gaps by isolating specific mechanism features across a larger set of chairs, and by incorporating brief adjustment-reasoning interviews directly into ease-of-adjustment protocols like Experiment 1's.

## 5. Conclusion

This study compared two commercially available, well-regarded office chairs, the T90 and the Aeron, to examine not only whether one outperformed the other, but why. Across two experiments spanning deliberate adjustment and dynamic, multi-device work, Chair A (T90) was rated as easier to adjust overall and received higher satisfaction for two of five components (exploratory, uncorrected comparisons), while Chair B (Aeron) was rated easier to adjust for one control. Rather than treating these results as a verdict on two specific products, this study used the chairs' documented mechanism differences and participants' interview responses to suggest the interface properties that may underlie each outcome: adjustment mechanisms that are physically accessible and operable in a single intuitive motion (as in Chair A's lumbar-support levers and armrest height button) were generally associated with lower perceived adjustment difficulty (with an exception for armrest fore-aft position, where Chair B's continuously sliding control was easier to discover), while mechanisms requiring indirect or high-effort operation (as in Chair B's rear-mounted lumbar dial and lever-based armrest height) were associated with the opposite. The one exception, seat height, where Chair B was rated easier despite both chairs using the same gas-lift mechanism, further suggests that adjustment difficulty can stem from how visible and recognizable a control is, not only from the mechanical operation it performs once located.

Equally informative were the findings that did not reach significance. For the headrest, armrest, and backrest recline, neither chair was rated significantly more satisfying than the other during dynamic use, a pattern that interview comments suggest may reflect each design leaving a different dimension of the same user need unaddressed, although limited statistical power cannot be ruled out. The armrest findings illustrate this most clearly: Chair A's armrest was significantly easier to adjust than Chair B's, yet this advantage did not translate into higher satisfaction during the movement task, possibly because ease of adjustment and stability once adjusted are separate design properties, as two participants reported that Chair A's armrest shifted out of position during use. This distinction, between designing a control to be easy to set and designing it to remain fixed under the reaching and reorientation that dynamic work requires, is a distinction suggested by this study's interview data, and one that a product-level comparison alone would not have revealed.

Taken together, these findings suggest that evaluating adjustability by range of motion or number of adjustable controls alone is insufficient for anticipating how a chair will actually be used. The same control can succeed on one dimension (ease of adjustment) and fail on another (postural stability, or support that tracks the body as it moves) depending on specific, identifiable mechanism properties. For chair designers, this points to concrete, transferable guidance: adjustment mechanisms should be physically accessible and operable in a small number of intuitive steps; once adjusted, controls such as armrests should hold their position under repeated movement rather than only under static load; and supports such as the headrest and lumbar region should engage the body across the range of postures used during real work, not only at the extremes of recline. More broadly, this study demonstrates a methodology (pairing controlled ease-of-adjustment testing with a dynamic, multi-device task and structured interviews) that can be applied to other chairs and adjustable furniture to generate this kind of design-level evidence, rather than relying on marketed adjustability alone.

This work is limited by its focus on two specific chairs rather than a systematic manipulation of individual mechanism features, and by its sample size (N = 20), both of which limit the extent to which the proposed design explanations can be generalized or treated as causally established. Future research should test the specific mechanism properties identified here (accessible single-motion adjustment, post-adjustment stability, and posture-tracking support) in controlled comparisons across a larger and more diverse set of chairs and users.

## Declaration of generative AI and AI-assisted technologies in the manuscript preparation process

During the preparation of this work the author(s) used Claude (Sonnet 5, Anthropic) solely for language editing to improve clarity, grammar, and readability. After using this tool/service, the author(s) reviewed and edited the content as needed and take(s) full responsibility for the content of the published article.

## CrediT authorship contribution statement

Yejin Kang: Conceptualization, Data curation, Formal analysis, Investigation, Methodology, Validation, Writing – original draft, Writing – review & editing.

Dusan Baek: Data curation, Formal analysis

Dongwoo Park: Data curation, Formal analysis

Kitae Hwang: Data curation, Formal analysis

Woojin Park: Data curation, Methodology, Supervision, Validation, Writing – original draft, Writing – review & editing.

## Data statement

Data will be made available on request.

## Funding

This study was supported by Sidiz, the manufacturer of the T90 (Chair A) evaluated in this study.

## Acknowledgments

None.

# Appendix A. Controls of Each Chair

| | Name of controls | Interface for Chair A | Interface for Chair B |
|---|---|---|---|
| 1 | Seat height | 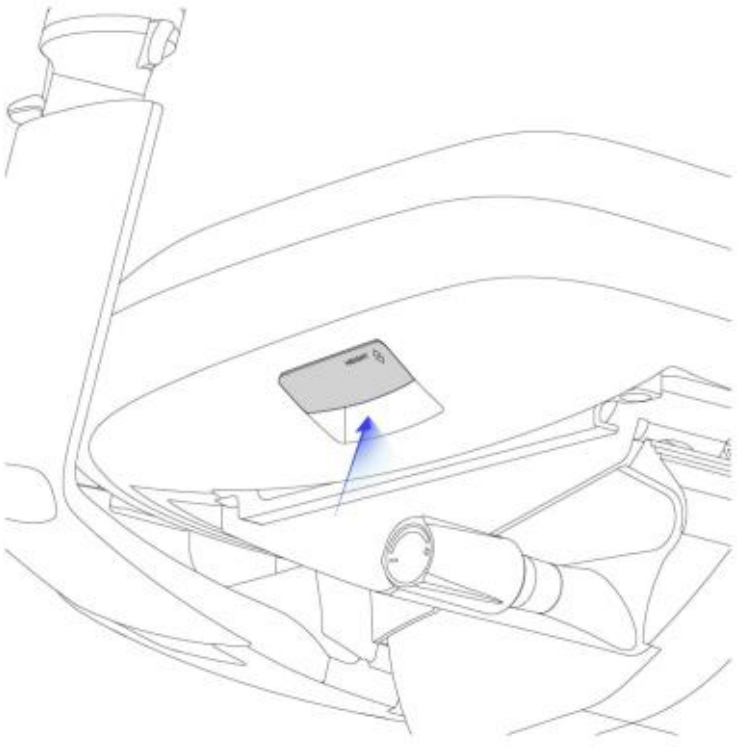 | 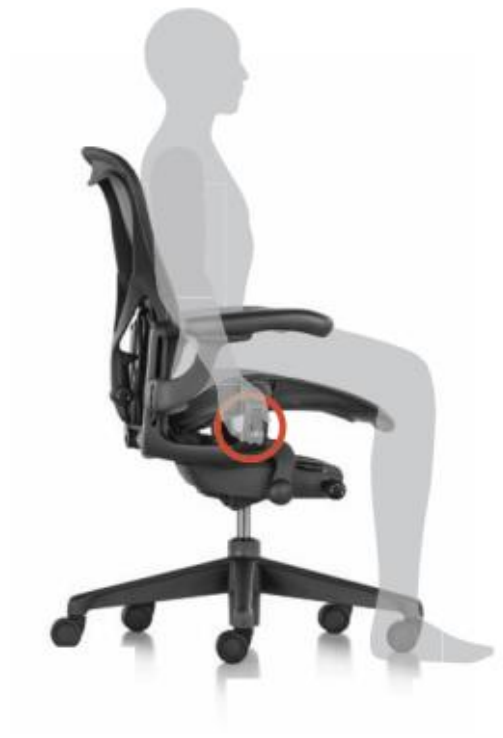 |
| 2 | Seat depth | 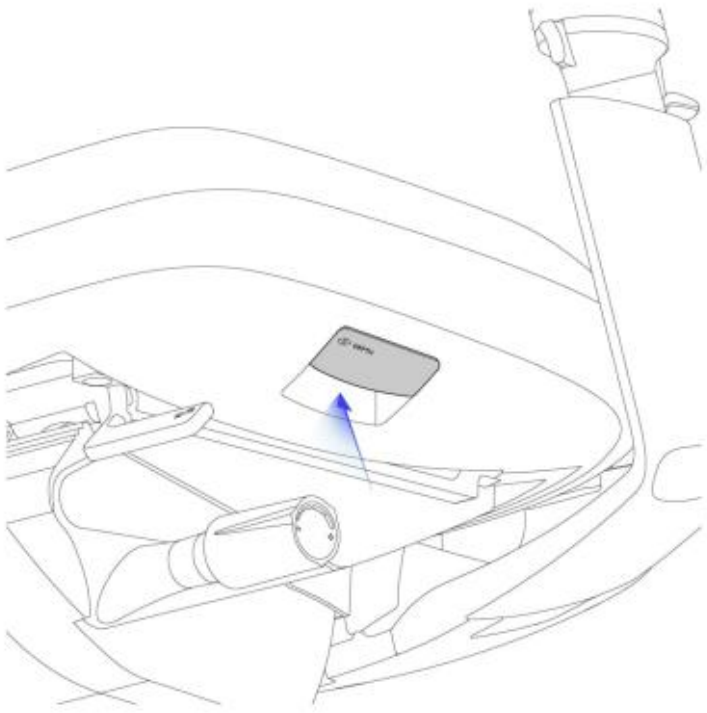 | None |
| 3 | Seat-pan tilt | 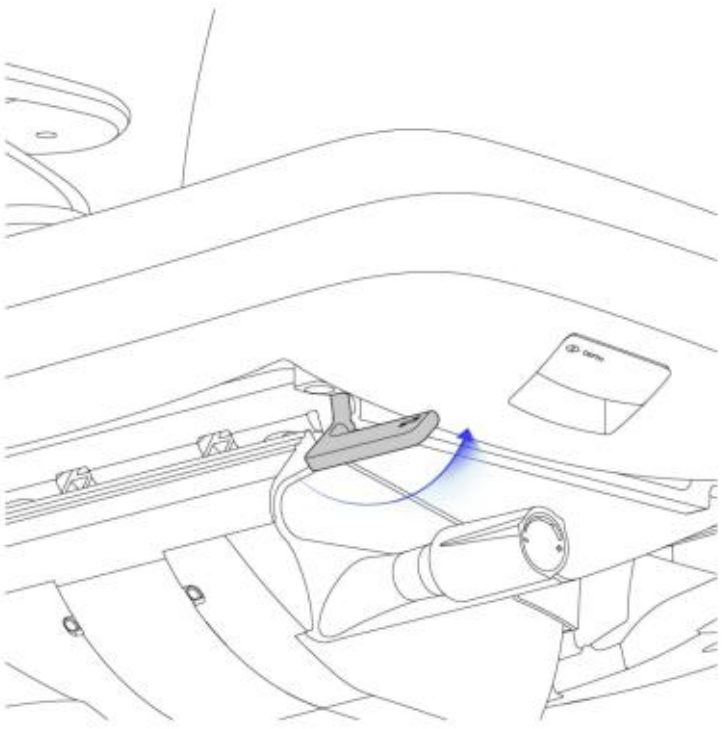 | 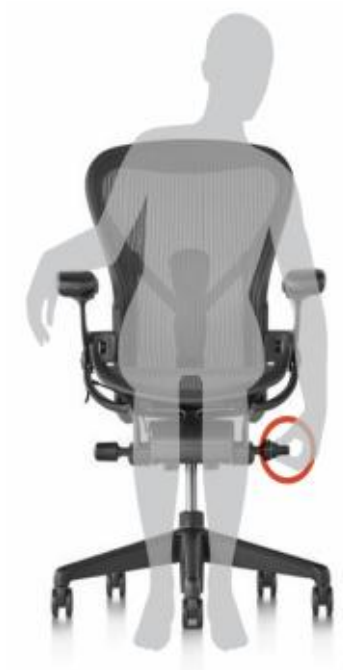 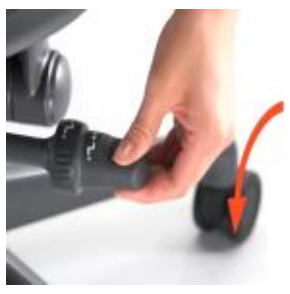 |
| 4 | Headrest height | 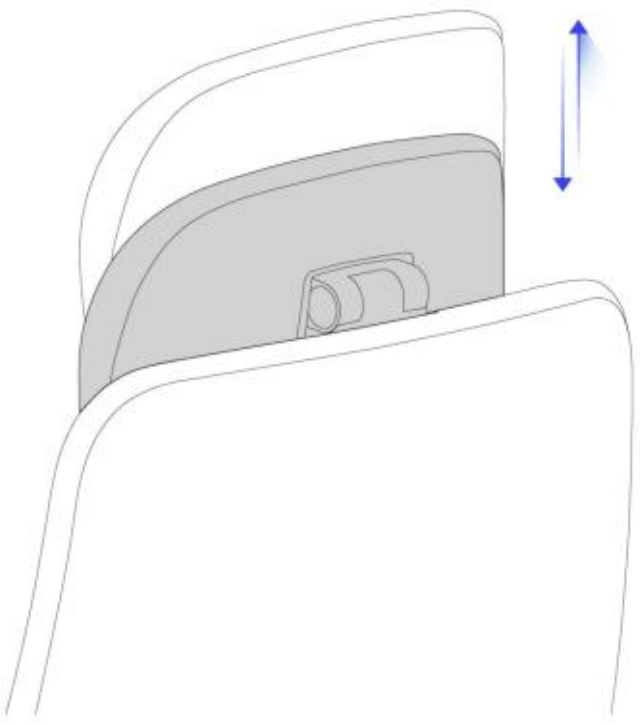 | 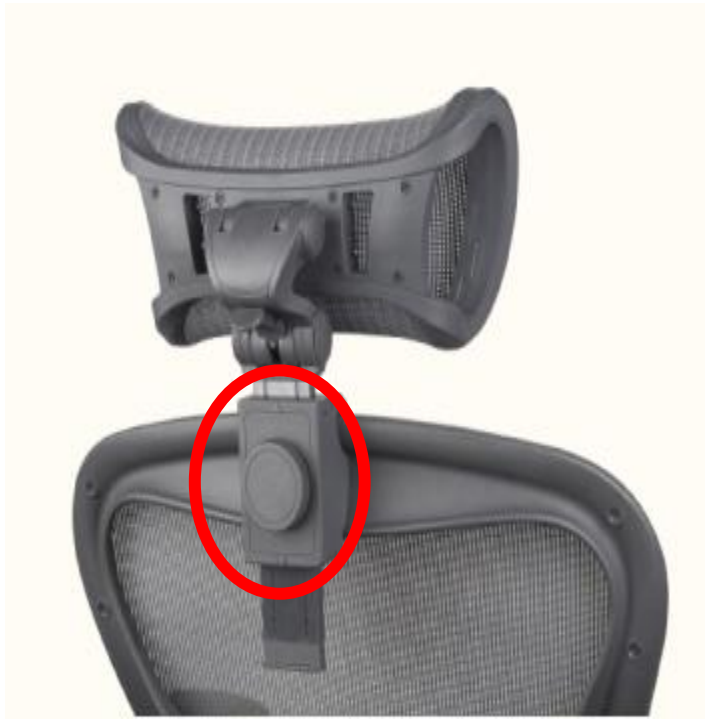 |
| 5 | Headrest angle | 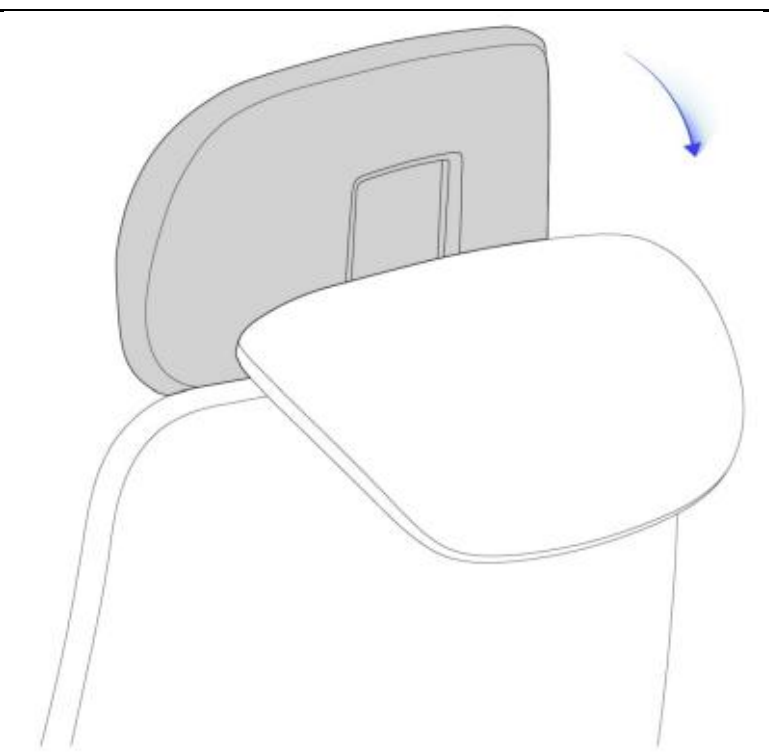 | 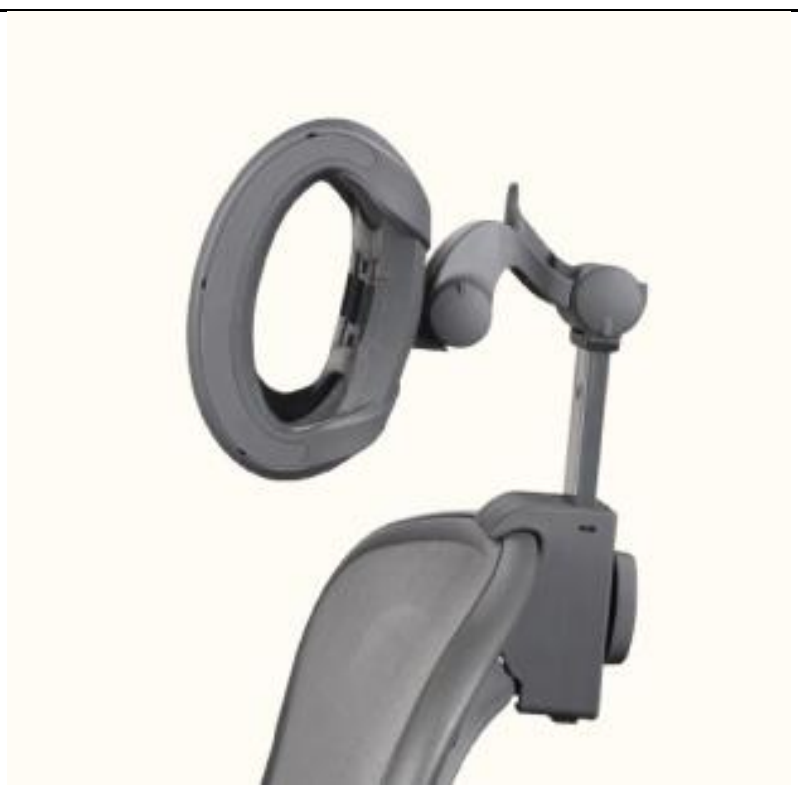 |
| 6 | Armrest height | 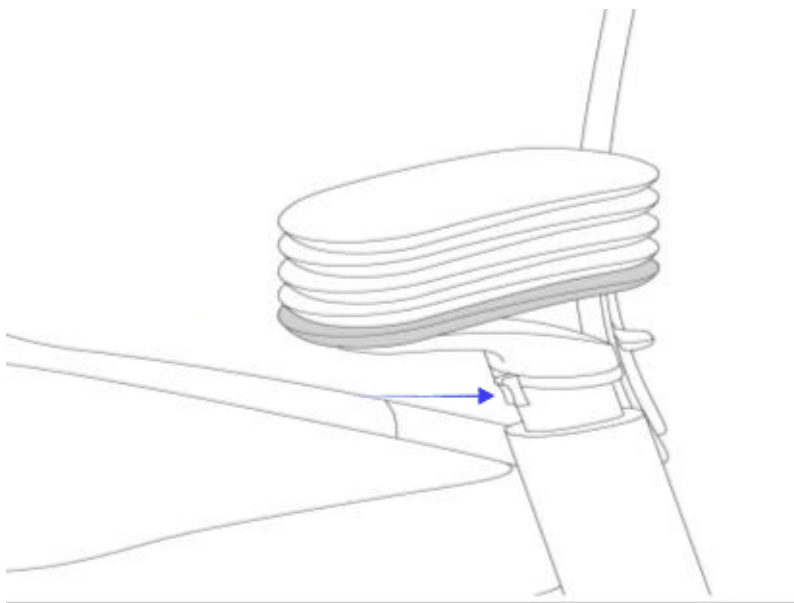 | 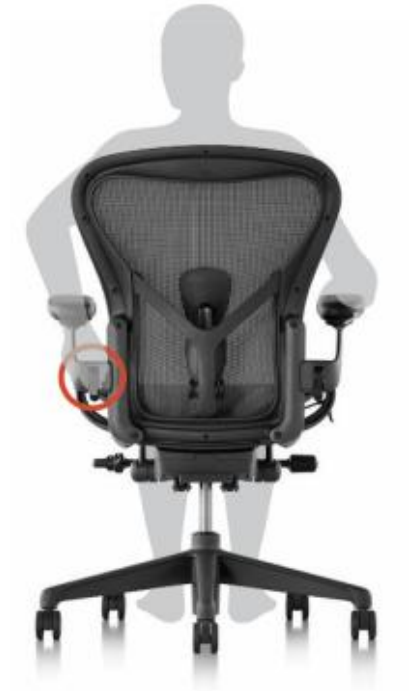 |
| 7 | Armrest rotation angle | 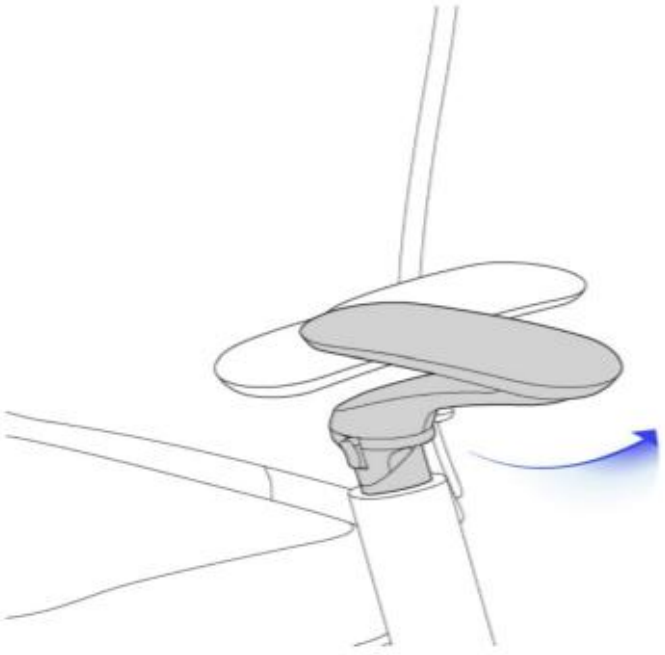 | 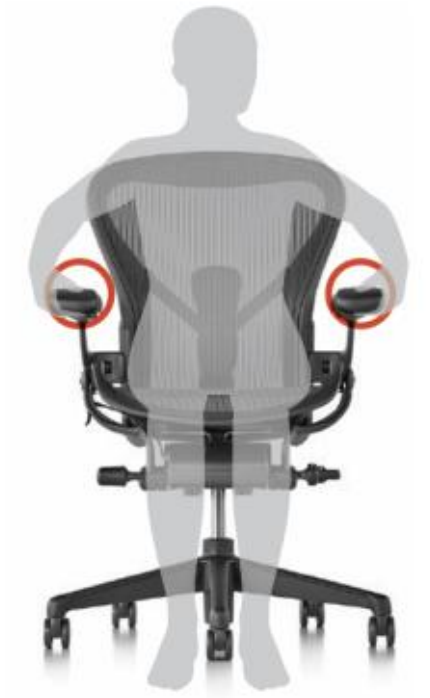 |
| 8 | Armrest fore-aft position | 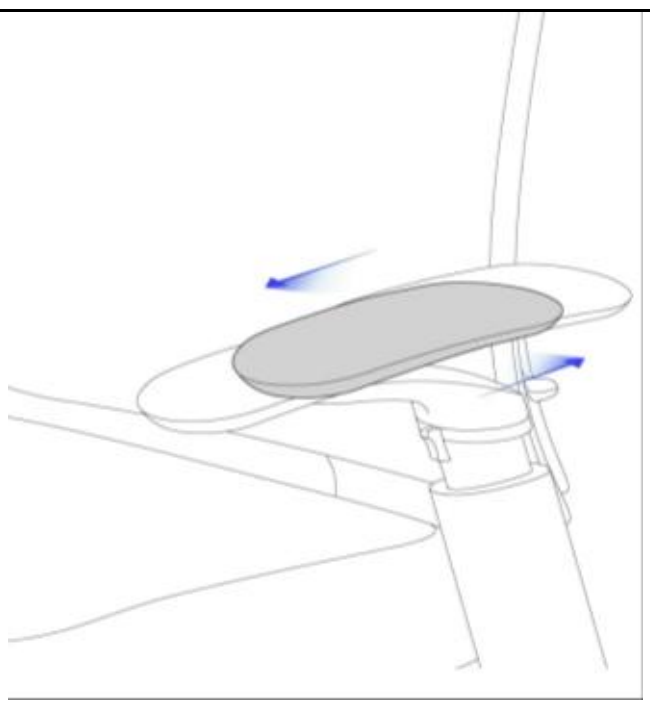 | 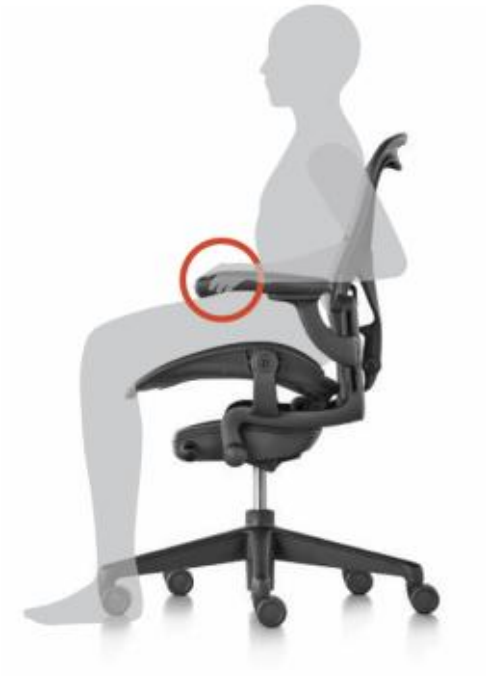 |
| 9 | Backrest recline tension | 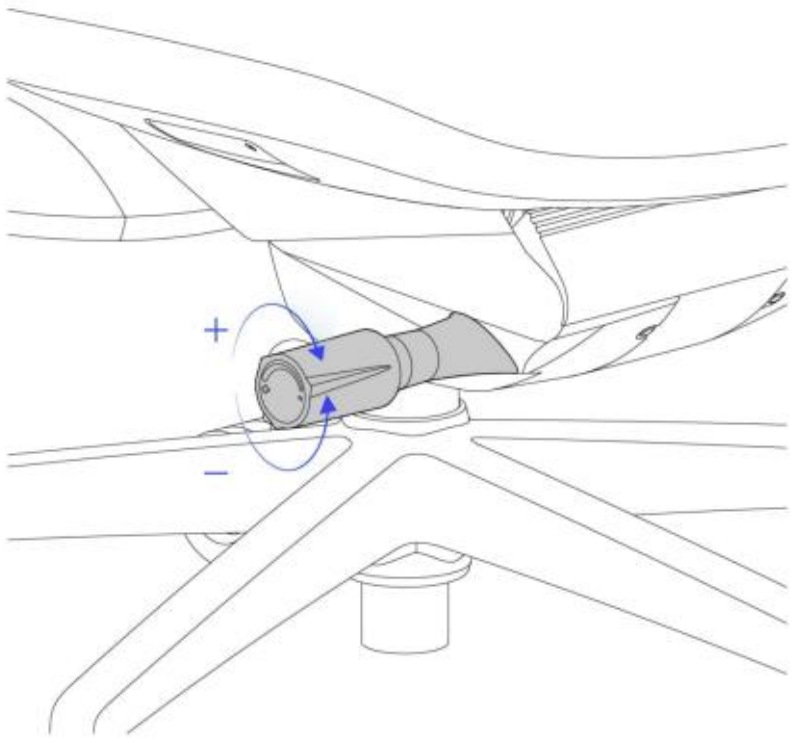 | 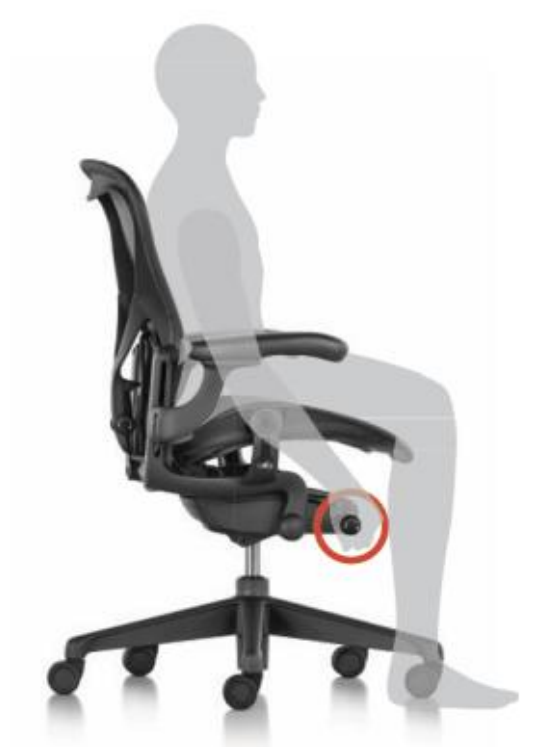 |
| 10 | Backrest recline limit | 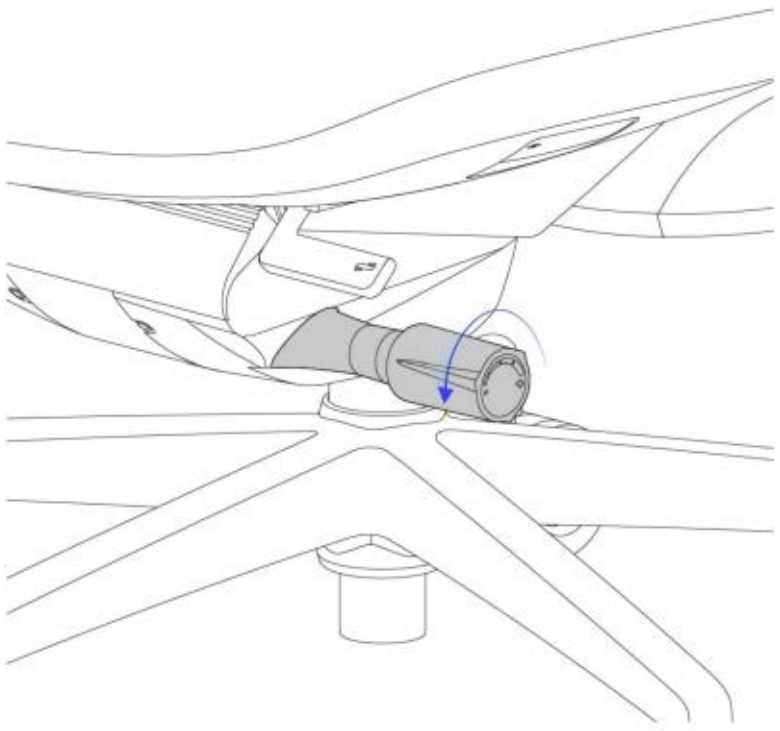 | 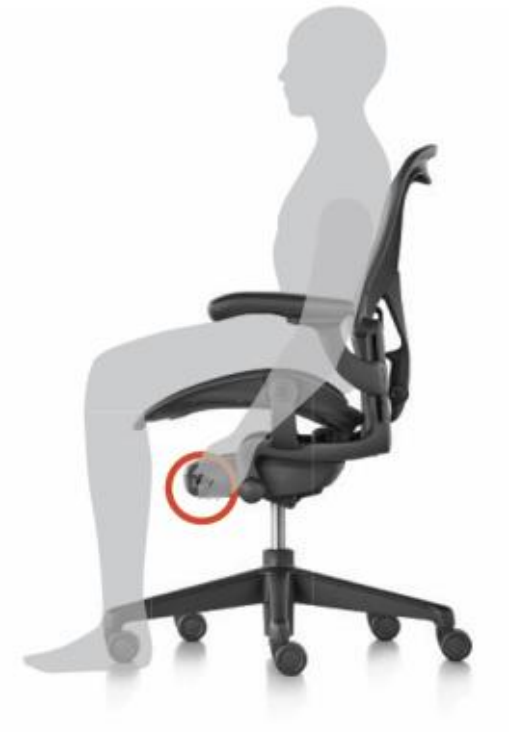 |
| 11 | Lumbar support position | 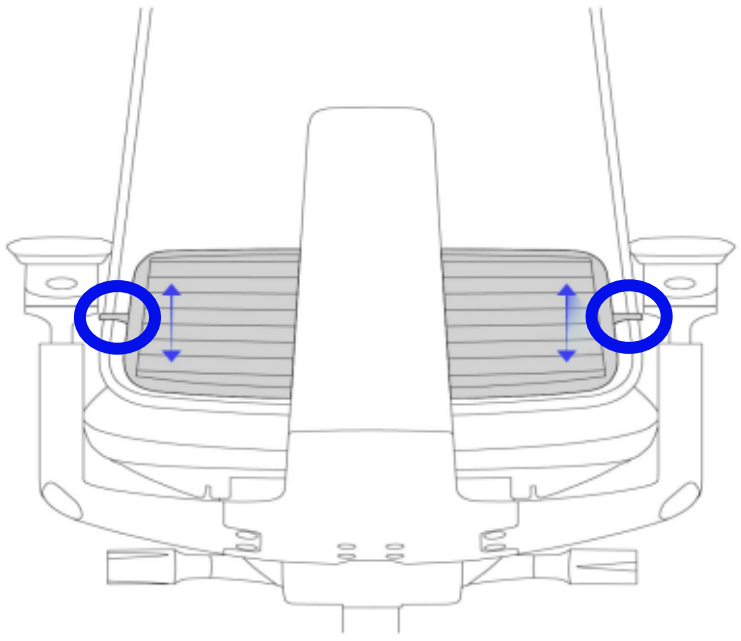 | 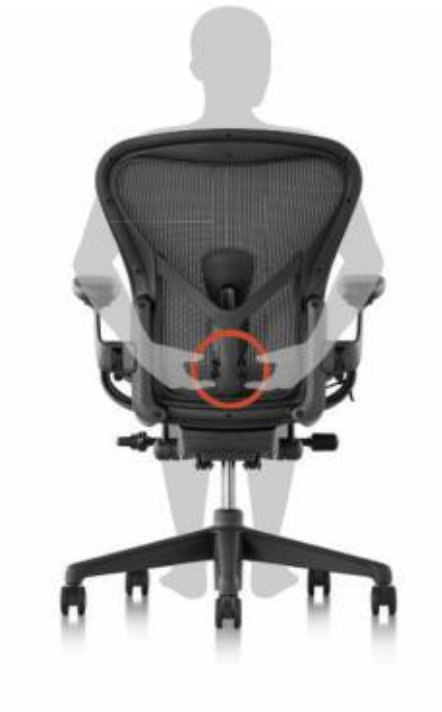 |